\documentstyle[aps,prb]{revtex}
\begin{document}
\draft
\title
{Second harmonic generation in SiC polytypes}
\author{Sergey N.~Rashkeev,\cite{a} Walter R. L. Lambrecht and
Benjamin Segall}
\address{Department of Physics, Case Western Reserve University, Cleveland,
OH 44106-7079}
\date{\today}
\maketitle
%\narrowtext

\begin{abstract}
A first-principles study of the  frequency dependent 
second harmonic generation (SHG) coefficients
of various SiC polytypes (2H, 4H, 15R, 6H, and 3C),
a group spanning the complete range of ``hexagonality'',
was carried out.  It uses a recently developed computational approach
based on the self-consistent linear muffin-tin orbitals (LMTO) 
band structure method, which is applied using the local density
approximation to density functional theory with a simple {\em a-posteriori}
gap correction. The susceptibilies are obtained 
in the independent particle approximation, i.e. without local field 
effects.
The zero frequency limits of the ratio $\chi^{(2)}_{333}/\chi^{(2)}_{311}$
for the noncubic polytypes were found to be in excellent agreement
to those obtained by the pseudopotential method -- and in disagreement
with simple geometric  predictions -- while the magnitudes
of the individual components themselves were found
to be smaller than the earlier calculated values.
The spectral features of 
the full $\chi^{(2)}(-2\omega,\omega,\omega)$ 
for 2H are  found to differ markedly 
from those of the other polytypes. The spectra in the series 
of decreasing degree of hexagonality, 4H, 15R, 6H 
gradually approach those for the zincblende (3C) form.
The independent tensorial components appearing in the rhombohedral
but not in the hexagonal forms are found to be about a factor 6 smaller
than the other ones.
An analysis of the SHG spectra in terms of $\omega$ and $2\omega$ resonances 
and individual band-to-band contributions is presented. 
It is suggested that second harmonic generation spectra 
have an advantage over linear optical spectra
for  probing the electronic structure, particularly for the region within
a few eV of the  band edges in that  they exhibit 
more detailed fine structure. That 
results from the sign variations in the products of matrix elements
occuring in the SHG.  
\end{abstract}


\section{Introduction} \label{sec:int}
Recently, there has been a renewed interest in calculations of 
second harmonic generation (SHG) and related nonlinear optical spectroscopies 
in semiconductors.\cite{Our1,Hughes,Hughes1,Allan,Chen1,Chen2}
In a previous paper,\cite{Our1} we presented our computational approach,
which is based on the linear muffin-tin orbital band structure method
and the recent formulations of the problem of evaluating the 
second order optical response functions for periodic solids 
in  the independent electron 
approximation by Sipe and Ghahramani\cite{Sipe} and Aversa and 
Sipe.\cite{Aversa} The independent particle approximation band structures 
used in this work are based on the density functional theory
in the local density approximation (LDA) with some 
{\em a-posteriori} self-energy corrections to the gap. 

In the course of our own  and other's 
previous work,\cite{Our1,Hughes,Hughes1,Allan,Chen1,Chen2} 
a question that has attracted some interest is the influence of 
the crystal structure on the second harmonic generation coefficients, 
in particular for materials which exhibit both the zincblende
and the wurtzite  structures.
Of course, the number of independent elements 
and the relations between the nonvanishing 
elements of the second order susceptibility
tensor depend on the symmetry (cubic or hexagonal) 
of the crystal structure. In addition, on the 
basis of  the strong similarity in the bonding, which is tetrahedral in both 
cases, one might expect there to be simple relations between the two tensorial
components $\chi^h_{333}$ and $\chi^h_{113}$ in the hexagonal structure 
and the one nonvanishing component $\chi^c_{123}$ in the cubic structure,
particularly in the static limit. 
These tensor relations are obtained simply by rotating 
the coordinate axes of the cubic 
coordinate system towards the hexagonal axes ($z$ along [111], and $x$ and $y$ 
chosen in the \{111\} plane such that the hexagonal rings
characteristic of both crystal structures are  aligned), 
and are given by:
\begin{equation}
\chi^h_{333}=-2\chi^h_{113}=2\chi^c_{123}/\sqrt{3}.
\end{equation}
Analogous relations have been used earlier for elastic 
constants.\cite{Martin,Kim}
While these simple geometric relations hold for the bond-orbital 
picture,\cite{Levine1}
in which the total polarizability is a sum of local bond polarizabilities,
they are not always well satisfied, as the near vanishing ratio of 
$\chi^h_{113}/\chi^h_{333}$  in wurtzite AlN 
demonstrates.\cite{Our1,Hughes1,Chen2}
The response functions of concern  describe the field dependent
polarizabilities of the electronic systems. These in turn are known to 
depend on the electronic structure which differ to a non-trivial
extent in the different polytypes (e.g. they have different band gaps).
Furthermore, the relation between SHG and the band structures is also
far from trivial and involves fine details of the electronic structures,
such as interband energy difference resonances and momentum matrix
element products.

Silicon carbide would appear to be the material of choice for a further
investigation of this question because of the occurence of 
polytypes,\cite{Verma}
which are structures of varying degrees of ``hexagonality''.
A detailed review of recent progress in the understanding of
the electronic structure and physical properties of the SiC polytypes
can be found in Ref.\onlinecite{Lambrecht1} and other papers 
in the same volume.
Specifically, ``hexagonality'' is defined 
as  $H=h/(h+c)$, where $h$ and $c$ are respectively
the number of hexagonal and cubic stackings in the elementary cell,
a parameter which varies from zero for the cubic 3C form to unity
for the purely hexagonal 2H polytype. Some other polytypes which
are frequently encountered, either because of their natural 
abundance or because their growth has been optimized are 6H ($H=1/3$)
and 4H ($H=1/2$).
There exist also rhombohedral polytypes such as 15R ($H=2/5$). These have
only threefold symmetry and thus have additional nonvanishing 
tensorial components, but we can still assign a hexagonality $H$ to them 
as defined above.  

In the static limit, Kleinman \cite{Kleinman} 
``permutation'' symmetry  allows one to reduce the number of independent 
tensorial components beyond the relations dictated purely by 
crystallographic symmetry. Thus the 333 and 311 components
fully describe the SHG susceptibility in the static limit.
The ratio of $\chi^h_{311}/\chi^h_{333}$ in the static
limit was studied previously for SiC polytypes by Chen et al..\cite{Chen1}
Additional  questions arise about: 1) the magnitude of the nonvanishing 
tensor components of the rhombohedral polytypes, 2)   the validity of 
these relations for the frequency dependent susceptibilities, 
including the additional independent tensorial components.

Some of these issues are investigated in this paper.
Although we are at present unable to explain the magnitudes of the
discrepancies from the expected geometric relations 
in simple terms, we can provide information 
on the general behavior of the frequency dependent 
$\chi^{(2)}(-2\omega,\omega,\omega)$ susceptibility tensor 
as function of $H$.
To  this end we performed calculations of  various tensor components of 
$\chi^{(2)}(-2\omega,\omega,\omega)$ 
for the 3C, 6H, 15R, 4H, and 2H SiC polytypes
and we relate them to the corresponding linear response 
function (more precisely to the imaginary part of the 
dielectric function, $\varepsilon_2(\omega)$)  and the 
calculated electronic band structures.
The chosen group of polytypes encompasses the complete range of
hexagonality. 

The relation of SHG to the band structures  
is an important question in its own right. In fact, one may
think of SiC polytypes as ``twist'' superlattices.\cite{Ikonic}
They consist essentially of narrow layers of cubic material bonded 
together with a 180$^\circ$ twist at the twin boundaries corresponding
to the hexagonally stacked layers. The result of these twists is a frustration
of the electron wave propagation.\cite{Backes} 
This leads to standing wave patters
which constitute essentially the ``mini-band-structure'' of the superlattice.
In particular, the   conduction band structure  
near the band edges can be thought 
of in this manner and show minigaps between folded cubic bands.
Just as nonlinear optics is currently of great interest in 
more conventional compositionally modulated 
semiconductor superlattices (in particular those with polar interfaces), 
it is  also expected to be of interest in the present context. 
Spefically, we will
show that the second harmonic generation spectrum 
in the range below the band gap but above half the band gap contains 
detailed information about the band structure. Measurements of these spectra 
in conjunction with the analysis presented here could thus provide 
a new experimental window on these aspects of the  polytype band structures. 
In that sense, the present work is closely related to our 
previous work on linear optical response functions,
\cite{Lambrecht1,Lambrecht2} which was also mainly  concerned with 
the question of extracting as detailed as possible experimental information
about the band structures from the optical data.

The SHG spectra  appear to be more sensitive to
the degree of hexagonality and to the underlying band structure 
than the corresponding linear response functions.
There are two main reasons for that sensitivity. For one thing,
the second order response involves more ``resonances'' than the linear
one. In addition to the usual $\omega$- resonances there appear 
the 2$\omega$- resonant contributions. Secondly, 
the real and imaginary part of the products of matrix elements
which controls the strength of a given resonance 
in $\chi^{(2)}$ can be positive or negative.
In contrast, for the linear responses the corresponding factors involve
only the square of matrix elements which insures, for
example, that $\varepsilon_2(\omega)$ is positive.
As a result, the structure in $\chi^{(2)}(-2\omega,\omega,\omega)$ is more
pronounced than in the linear response.
Hence, measurements of the frequency dependent nonlinear responses
can in principle provide more detailed information about the electronic
structure than those for linear response.
In that sense these measurements are similar to modulation spectroscopy 
techniques.\cite{Cardona}
Another factor favoring the nonlinear studies is that the threshold
for the 2$\omega$- part occurs at half the energy of threshold
for linear processes. The measurements could thus
be performed in a more convenient spectral region for the incident 
light. Although one still needs detection capabilities
of the second harmonic in the doubled frequency range, at least the 
intensity requirements for the incident light are thereby somewhat relaxed.

Unfortunately, at this point there has been only a very limited
amount of experimental work on SHG in SiC in spite of  
the technological importance of this material.
So far, there were measurements
of the SHG only for the zero frequency limit;
and those few were carried out 
more than two decades ago.\cite{Singh,Miller}
Very recently, Niedermeier et al. \cite{Niedermeier} 
carried out measurements of the anisotropy 
of the SHG on 3C SiC films
on Si substrates and 6H and 15R crystals  grown by the modified Lely 
method.  There is thus some renewed 
experimental interest in such measurements. At present, 
the primary interest of this work  appears to be characterization 
of the quality of SiC samples. We hope that our present work will 
stimulate further interest in  the more fundamental questions 
addressed here.
In view of the absence of experimental studies of the
frequency-dependent SHG,
our first principle calculations of the SHG have a predictive
character.

The rest of the 
paper is organized as follows. In Section \ref{sec:comp} we discuss
our computational approach based on the LMTO band structure code.
In Section \ref{sec:res} we analyze the results of calculations
and find the main similarities and differences between the
nonlinear responses for different SiC polytypes.
A conclusion and summary of the results are presented in Sec. ~\ref{sec:con}.

\section{Computational Method} \label{sec:comp}

The expressions used in the present work to calculate $\chi^{(2)}$
were given in a previous publication.\cite{Our1} They are a rearrangement
of the formalism obtained by
Sipe and Ghahramani,\cite{Sipe} and Aversa and Sipe\cite{Aversa}
using the ``length-gauge'' formalism.
Those results are based on the independent particle
approximation (meaning that no local field effects are included)
for undoped semiconductors, i.e. systems without partially filled bands.
There are  several advantages in using this formulation,
namely: (i) the manifest absence of unphysical 
singularities in the zero-frequency
limit; (ii) the simple and natural account of the effects
of the intraband motion of electrons which give an essential
contribution to SHG; (iii) the obvious satisfaction of the
Kleinman relations\cite{Kleinman}
in the zero-frequency limit.

This formulation has recently been used successfully 
in two sets of studies. In one set Hughes et al. calculated 
$\chi^{(2)}(-2\omega,\omega,\omega)$ in GaAs and GaP \cite{Hughes}
which have the zincblende structure,
and in GaN and AlN \cite{Hughes1} which form with the wurtzite structure.
These calculations were based on 
band energies and momentum matrix elements obtained
by the self-consistent full-potential
linearized augmented plane wave (FLAPW) band structure method.
The other set was carried out by the present authors \cite{Our1}
who studied 3C-SiC in addition to the above-mentioned four
semiconductors.
In those calculations we used 
the self-consistent 
linear muffin-tin orbitals method (LMTO)
\cite{OKA,lmto} within the atomic sphere approximation (ASA).
The same approach will be used here. The method is very efficient
mainly because it employs a rather small basis set.
As a result it can more easily
deal with systems containing a large number of atoms per unit cell 
while maintaining a sufficiently large number of 
{\bf k}-points so as to insure converged Brillouin zone (BZ) integrations.
Extensive checks performed in our 
previous paper \cite{Our1} demonstrated that our LMTO based approach
yields accurate results for the second
order response functions.

The self-consistent calculations of the electronic band structure
(eigenvalues and eigenstates) were carried
out within the framework of density functional theory in the local
density approximation (LDA) \cite{ldft}
using the exchange-correlation parametrization of
Hedin and Lundquist.\cite{Hedin}
As was extensively discussed in 
our previous paper,\cite{Our1} correcting for the well-known ``gap problem''
of LDA is extremely important for nonlinear response functions.
While this is true even in the static limit, as can be justified from 
the point of view of the recently developed concept of 
polarization$+$density functional theory (PDFT),\cite{Aulbur} it is clearly 
imperative for frequency dependent response functions to somehow 
deal with the actual quasiparticle excitations rather than the 
Kohn-Sham eigenvalues. 
One of the most accurate approaches presently available for the 
corrections to the LDA is  the
GW approximation.\cite{HedinGW} However, calculation of the full 
energy dependent and nonlocal self-energy operator 
even in this first relatively simple approximation 
is rather cumbersome in practice because of the 
need for determining the fully dynamically 
screened Coulomb interaction $W$. 
A simplified approach is based on the observation that the 
conduction bands to a good approximation shift up rigidly in GW 
calculations. The effect can thus be described by a 
the so-called ``scissors operator'', which can be written as 
a projection operator on the conduction bands times a constant shift
in energy $\Delta$. As was pointed out by  
Levine and Allan \cite{Allan} and later by Hughes and Sipe,\cite{Hughes} 
the introduction of this shift operator into the Hamiltonian results in
renormalization of the velocity (momentum) operator matrix elements.
In practice, this renormalization factor is taken simply as 
\begin{equation}
{\bf p}_{nm}\mapsto{\bf p}_{nm}\frac{\omega_{nm}+
(\Delta/\hbar)(\delta_{nc}-\delta_{mc})}{\omega_{nm}}, \label{pshift}
\end{equation}
in which $p_{nm}$ is a momentum matrix element between Bloch states
$n$ and $m$, $\hbar\omega_{nm}=E_n-E_m$ is their band difference,
and the factor $(\delta_{nc}-\delta_{mc})$ limits the 
corrections to matrix elements involving one valence and one conduction band.
That approximation is based on the explicit assumption that the 
dipole moment $r_{nm}$ matrix
elements are unchanged because the perturbed wave functions are close the LDA 
wave functions. In our previous paper, \cite{Our1} we noted some 
principle shortcomings of this approach  in that it breaks the consistency
between the eigenvalues and eigenvectors. We showed that better results
were obtained by introducing the shift at the level of the LMTO 
Hamiltonian by an empirical modification of diagonal elements 
corresponding to the basis functions that primarily make up the 
conduction-band states. These typically  are the cation and empty sphere
s-like states for the lowest conduction bands in 
zinblende semiconductors.  However, at present this approach only allows us to 
shift the lowest conduction bands rather than the whole set of 
conduction bands which appears to be more appropriate for SiC. 
Also, a fine tuning of this approach for SiC polytypes which are described
using different empty spheres for the hexagonal and cubic local stackings 
remains to be carried out. In the interest of simplicity 
and a consistent treatment of all polytypes, we therefore 
here adopt the ``scissor's approach'' as described above. 

The next question is what value  of the constant shift $\Delta$ to adopt.
The ``scissors'' approach gives a reasonable quantitative 
agreement with experiment
for a variety of optical constants for moderately small band gap
semiconductors when the value of $\Delta$ is straightforwardly 
taken to the difference between the experimental and LDA minimum band gaps.
However, as was noted by Chen et al. \cite{Chen1} 
and by Gavrilenko et al. \cite{Gavrilenko} the shifts required to
reproduce the magnitude of the experimental dielectric constants 
$\varepsilon_1(0)$
for the SiC polytypes are less
than the $\Delta$ needed to match the band gaps. 
Similar results occur for other large band gap semiconductors
which contain second-row elements
(B,C,N,O), i.e. the discrepancy in the LDA optical response functions 
are overcorrected by the use of the ``scissors'' 
approach.\cite{Chen1,Chen2,Aulbur} 
In our opinion, this problem is further complicated by the effects 
of local field corrections and continuum excitonic effects on the 
oscillator strength. Therefore, we think it is premature to attach
too much importance to the magnitudes of the response functions,
especially since we are presently not including local-field corrections.
We prefer to focus on the energetic position of the spectral features. 
It was shown that a simple self-energy correction of about 1 eV for all the
major SiC polytypes appears to bring the LDA values for the minimum gaps 
into good coincidence with experiment.\cite{Lambrecht1}
This value is well justified by recent calculations of such corrections
using the GW approximation.\cite{Rohlfing,BackesGW,Wenzien}
Also, a constant energy shift   of $\varepsilon_2(\omega)$ 
by 1 eV leads to reflectivities
in a good agreement with
measured reflectivity spectra \cite{Lambrecht2} as far as 
the location of spectral features is concerned. Since our major
interest for $\chi^{(2)}$ is also in the location of the spectral features, 
we therefore adopt the value $\Delta=1$ eV. From our 
experience with other materials, we expect that both the use of 
the scissor operator approach in its present form and the choice of 
$\Delta$ may somewhat overestimate the correction to the LDA. 
We will keep this in mind when comparing to other values in the static
limit. It should not affect our major conclusions about   the spectral 
features. 
Local-field correction were shown to be not larger than 10\%
for the zero-frequency limit of $\chi^{(2)}$ 
in the SiC polytypes.\cite{Chen1} 
They are neglected in the present calculations. 

Next, we turn to  some computational details.
The imaginary part of the frequency dependent
SHG is calculated first (see Ref. \onlinecite{Our1}).
The real part of the SHG
is then obtained from the Kramers-Kronig transformation.
However, in the 
zero-frequency limit, the SHG
can be evaluated with less effort by the use of
a special expression. 
A comparison of that result with the limit of the frequency dependent
$\chi^{(2)}$ serves as a check of the accuracy
of both the BZ integration and the Kramers-Kronig transformation.
We note that the {\bf k}- integration
can be limited to the irreducible wedge of the Brillouin zone
(IBZ) only 
if a preliminary symmetrization of the product of the
3 momentum matrix elements over all the transformation of the crystal group
is performed.
For the frequency dependent SHG, we use the usual tetrahedron scheme 
for the integrations with linear interpolation of the band energies and the
products of the matrix elements. 
For the zero-frequency limit on the other hand,
we employ a semi-analytical linear
interpolation scheme which is more efficient and produces a smaller 
error.\cite{Our1}

Orbitals with angular moments up to $l_{max}$=3 were included
in the basis set.
As shown in our previous work \cite{Our1}
neglecting the f-states leads to non-negligible errors
in the momentum matrix elements and the SHG's. 

We finish this section with a note on the symmetry allowed tensorial
components.
In the cubic  case, there is only one independent component, 
with indices 123,
and all its possible permutations are equal to it;
(1, 2, and 3 refer to the x, y, and z axes respectively,
which are chosen along the cubic axes).
In the hexagonal polytypes which correponds to the point group 6mm,
there are 3 independent components with indices 
333, 311, and 131, the last one equaling the 113 component because 
for SHG the last two indices can always be permuted.
The coordinate axes here are chosen with z along he 6-fold 
symmetry axis. Furthermore, in the static limit
the 311 and 131 components are equal by the Kleinman 
permutation symmetry but this is no longer true for the frequency 
dependent case.
The rhombohedral polytype 15R (point group 3m) has a threefold symmetry
axis along the z-axis, which is normal to the basal planes, instead
of the six-fold screw axis present in the hexagonal structure.
As a consequence of the lower symmetry there is one
additional independent component of the SHG, 
and several symmetry-related nonvanishing components. These are 
$\chi^{(2)}_{222}=-\chi^{(2)}_{211}=-\chi^{(2)}_{112}=-\chi^{(2)}_{121}$.
As noted in the introduction, a  transformation of the 3C
$\chi^{(2)}$ tensor from a cubic coordinate system
to a hexagonal one (having the z-axis along the cubic [111]
direction) yields the 333, 311 and 131
components in terms of the 123 response function
for 3C. 
The resulting ratio $\chi_{333}^{(2)} / \chi_{311}^{(2)} = -2 $ 
for arbitrary frequency $\omega$. The $-2$ value
found for the ratio  
when one uses directly computed components for 3C 
in the hexagonal coordinate system represents
an additional verification of the computer code. 
One of the central questions of this paper is to what extent the 
same relation holds for the other polytypes.

\section{Results} \label{sec:res}
\subsection{Static limit} \label{static}

The zero frequency limit of the SHG being simpler to calculate,
it is the first aspect of the nonlinear response on which we will 
focus our attention. To put our own results in perspective,
we briefly recall the history of static SHG results for SiC. 
The first such calculation for SiC was carried out 
by B. F. Levine using the bond charge model.\cite{Levine1}
He found that the  
$\chi_{123}^{(2)}$ in zincblende SiC and 
$\chi_{333}^{(2)}$ in wurtzite crystals are both negative.
Subsequent analysis of the bond-charge model found  
that its predicted SHG values in SiC
are very sensitive to the choice of the bond charge and the ion radii
because both Si and C have the same ionic charge.\cite{Singh}
However, the ratio of the 333 and 311 tensor components 
in this model is found to be independent
of the particular bond charge model and to be equal to the above mentioned 
value of $-2$ for all polytypes. For the 3C polytype,  
the ratio is purely geometric in origin. For this value to hold for the
other polytypes, the approximation implied is one of 
similarity in the underlying electronic structure.

The first indication that the real situation is more complicated,
and that the bond-charge model is not accurate at least
for some hexagonal
materials including nH-SiC, appeared in the work of 
Chen et al. \cite{Chen1,Chen2}
Our calculated  ratios of the 333 and 311 components of $\chi^{(2)}$
for the zero-frequency limit are shown as function of 
hexagonality in Fig.\ref{fig:ratio} along with 
Chen et al.'s \cite{Chen1} results.
It can be seen that the deviations for the non-cubic forms increase
with increasing hexagonality becoming substantial for the large
values of $H$.
Our values for the ratio are in excellent agreement with those obtained in
the LDA pseudopotential calculations of Chen et al.,\cite{Chen1}
which include local-field corrections.  
A comparison of the values of the individual 333 and 311
components obtained in the present and pseudopotential calculations
is presented in Table \ref{tab:chi0}.
It is seen that while there is good agreement for the ratios,
the absolute LDA values
of the $\chi^{(2)}$ appear to be more sensitive to the method of
calculation.
Those obtained by the LMTO method are found to be smaller than
those by the pseudopotential approach. 
As expected, the gap corrections reduce the values in both our 
and Chen et al.'s calculations. A slightly different value of the gap 
correction  was used for each polytype in their case, 
with $1.04<\Delta<1.27$ eV.

Although there are non-negligible differences between 
our and their $\chi^{(2)}$ values, even in the LDA, we wish to point 
out that the two results agree much better with each other than 
with the bond charge model predictions. The latter predict values 
independent of polytype which in Ref. \onlinecite{Levine1}
are  $\chi_{333}=-166$, $\chi_{311}=84$ pm/V. Singh et al. \cite{Singh}
pointed out that values about 10 times smaller
$\chi_{333}=-16$, $\chi_{311}=8$ pm/V are obtained when different 
ionic radii values are used as input to the same model. 
The two-band model prediction of Kleinman \cite{Kleinman1} gave  
$\chi_{333}=-80$, $\chi_{311}=40$ pm/V. 
Note that all of these have opposite sign to ours.
Experimentally, the values available correspond to an incompletely 
determined  polytype\cite{polynote} and are 
$\chi_{333}=\pm27\pm3$, $\chi_{311}=\mp15\pm2$ pm/V 
at $\lambda=1.064\,\mu{\rm m}$.
These values are obtained by rescaling the 
original values  by Singh et al.,\cite{Singh} which  were
relative to  quartz, using  a more recently obtained absolute value 
of quartz recommended by Roberts..\cite{Roberts} 
For the 311 component, the average was taken of 
the experimentally slightly different 311 and 113 components.  
As in the original measurement, we leave the absolute sign undetermined,
although the relative sign of the two components was clearly established.
By applying Miller's rule \cite{Miller64} that
$\chi_{ijk}^{(2)}(-2\omega,\omega,\omega)/[\chi_{ii}^{(1)}(2\omega)
\chi_{jj}^{(1)}(\omega)\chi_{kk}^{(1)}(\omega)]$ is independent of 
frequency in the low frequency regime, 
Chen et al. \cite{Chen1} converted these 
values to $\chi_{333}=\pm23\pm3$ and $\chi_{311}=\mp14\pm2$ pm/V in the 
zero frequency limit.
Our LDA values are apparently closer to these
values than the ``scissor'' corrected values. As noted 
earlier, we expect our ``scissor'' values to be overcorrected, 
both because of the intrinsic problem of the scissor's approach
and our use of the actual gap correction rather than a scaled down 
value which would reproduce the magnitude of $\varepsilon_1(0)$. 

We note that the signs of all the components in Table \ref{tab:chi0} are 
opposite to those reported in Ref. \onlinecite{Chen1}. 
The reason for this reversal in the sign of the $\chi^{(2)}$
is simply that opposite  coordination of the atomic positions 
was employed in the two sets of calculations: In 3C, for example,
we placed the Si atom at the (0, 0, 0) and the C atom
at the (1/4, 1/4, 1/4) positions in the unit cell while in 
Ref. \onlinecite{Chen1} the positions of the Si and the C atoms
were interchanged.\cite{Chen_pc}
There were corresponding reversals for the
hexagonal polytypes.
The agreement in the absolute signs of the $\chi^{(2)}$
components in the two sets of SiC calculations noted above, along with
a corresponding agreements for several other semiconductors discussed
in a previous paper,\cite{Our1} gives us strong confidence
in that regard. However, it appears that the experimental data 
are still unclear about the sign
despite an early effort to determine the absolute sign of the 
SHG.\cite{Miller}  In that paper, a negative sign is proposed for $\chi_{333}$
based on a surface etching method for the determination 
of the crystal orientation. This also coincides with the 
reported sign of the bond-orbital methods, which however, could easily 
give  either sign depending on details of the model and is in 
disagreement with our sign.
An unambiguous measurement of the signs, as well as
the magnitudes, clearly would be useful.

\subsection{Frequency dependent results} \label{freqdep}
Figure \ref{fig:chi_pol} shows the 333, 311 and 113 components of
the $\chi^{(2)}(-2\omega,\omega,\omega)$ for the complete
set of SiC polytypes considered here
over the energy range of 0 to 8 eV. 
We note that the hexagonality parameter $H$ decreases monotonically 
from 2H ($H$=1), shown at the top of the figure,
to the 3C ($\infty$H) polytype ($H$=0) at the bottom.
As mentioned above, the ratio of the 333 to the 311 component
for 3C is the fixed geometric value of $-2$. For all other
polytypes the ratio is highly frequency dependent
in the absorbing frequency regime, i.e. for $\hbar\omega>E_g^d/2$,
with $E_g^d$ the smallest direct gap. 
Below, this frequency, the ratio is remarkably constant as can 
be seen in Fig. \ref{figratiofreq}.
Even in the absorbing frequency range,
the ratio of the coefficients averaged
over a modest energy of say 0.5 to 1 eV are roughly in $-1$ to $-2$
range with the deviation largest for the largest $H$, that is, for
the wurtzite structure. 

It is striking that the 333 component of the wurtzite (2H) 
structure is very different from all other polytypes 
in that the first peak (at about 4 eV) is negative and the second 
one positive. In all other polytypes, the first peak
is positive and the second negative, with the location of the 
first peak gradually shifting towards the position of 3C.
In fact, one can see that 
the spectra for 4H, 15R and 6H show roughly a superposition 
of the 2H and 3C characteristic shapes: i.e. they have a $+-+-$
sequency of peaks. One may interpret the first positive peak
as cubic-like, the next negative and positive peaks as hexagonal like
and finally the negative tail appoaching zero as cubic-like.
This is not unexpected because the polytypes
can be viewed as mixtures of locally cubic and hexagonal
stackings. Of course, one cannot push this model too far
because the band structures from which the spectra are derived 
are not simply a superposition of local contributions but  
contain interference effects of the electron waves. Thus, 
the spectra are not simply a linear superposition in the appropriate 
ratios of cubic and hexagonal components
indicated by the hexagonality. 

In the   311 and 113 components,
2H appears to be more similar to the other polytypes. 
Considering the tensorial aspects,
one may note the close similarity in spectral features between 
the 311 and 131 components which are exactly equal in the static limit.

Also evident in the figures is the general
increase in the complexity of the spectral fine 
structure with decreasing $H$ (or,
increasing n in the hexagonal nH polytypes) until the cubic structure
is reached. This is most clear for the 333 component.
This fact clearly reflects the increasing folding 
of the bands with n or more generally with the size of the basic
repeat unit in the stacking sequence. The latter is 2 in 4H and 3 in 6H
because they correspond  to bands of two and three consecutive
cubically stacked layers alternated with a twin boundary (or hexagonal
stacking). In 15R the repeat unit in this sense is $\langle23\rangle$
although the periodic repeat unit along the c-direction is  15 layers.
This notation scheme based on the width of the 
cubically stacked  bands is known as the Zhdanov  notation.\cite{Zhdanov}
Of course, some of the fine structures 
exhibited in curves is expected to be averaged out in the observed spectra
as a result of broadening effects.  

Finally, we consider the additional independent component
for the rhombohedral structure.
The 222 and 211 components for the 15R polytype are displayed
over the range of 2 to 8 eV in Fig. \ref{fig:trig}. 
It is evident that these accurately satisfy the symmetry requirement
of having opposite signs. This represents just one more test of the 
correctness of the code. The 112 and 121 
components were also calculated and, as required, are found
to be virtually identical to the 211 component. Finally, we note
that the magnitude of these components tends  to be smaller than those
for the 333 and 311 components by a factor of about 6. These
small magnitudes reflect the similarity in the symmetry of
the rhombohedral and hexagonal structures.

\subsection{Analysis of SHG spectra}\label{analysis}
As we noted in a previous paper,\cite{Our1}
it is convenient to analyze
the $2\omega$- and $\omega$- resonant contributions in the SHG separately
as the two parts have features at different energies.
A comparison was made there between those two parts and
$\varepsilon_2(2\omega)$ and $\varepsilon_2(\omega)$ respectively
for 3C-SiC (see Fig. 8 of Ref.\onlinecite{Our1}).
As might be expected, the location of structures
in both parts of $\chi^{(2)}$ coincided with those in the dielectric 
function. It should be noted that the threshold for the 2$\omega$-
part occurs at an energy that is half of that for the $\omega$- part,
$E_g^d$, which is the direct transition threshold for
$\varepsilon_2(\omega)$. As a result, only the 2$\omega$- part
contributes to $\chi^{(2)}$ in the important energy range
$E_g^d/2 < \hbar\omega < E_g^d$. Since 
$E_g^d \geq$ 6 eV for
the SiC polytypes, unless one is concerned with fairly high energies
(well into the UV), the 2$\omega$- part is the one of primary
interest.

As in $\varepsilon_2(\omega)$,
the origin of structure in $\chi^{(2)}$ can be analyzed by a decomposition
into separate band-to-band
contributions (see, e.g., Ref.\onlinecite{Lambrecht2} for details).
The main two peaks in the $\varepsilon_2(\omega)$ curve for 3C-SiC,
for example, result from 
transitions between the upper valence band to the
first and second conduction bands respectively. Previous analysis 
shows that the peak with lower energy comes from extended regions
of {\bf k} space 
near the cubic $\Gamma$-K-L plane and close to the $\Gamma$-L line.
The transitions between the same bands and {\bf k}- 
space regions provide the contributions in the
2$\omega$- and $\omega$- terms of the $\chi^{(2)}(-2\omega,\omega,\omega)$.
These contributions to the total SHG (mostly to
its 2$\omega$- part in the energy region considered)
from these interband transitions
are displayed in the top panel 
of Fig. \ref{fig:par1} by the dotted and 
dashed curves. A similar analysis can be made for 2H, and, the
contributions from the relevant interband transitions are similarly indicated
in the lower panel of the figure.
The major difference between the dielectric function and 
$\chi^{(2)}$ could loosely be ascribed to the different matrix
element factors involved. In the former, only the absolute square of 
matrix elements occur insuring that $\varepsilon_2(\omega)$ be positive.
The situation is more complicated for the SHG. Since a complex product
of three generally different matrix elements appear in $\chi^{(2)}$,
its real and imaginary parts can assume either sign. This is evident in
the results for 3C and 2H shown in Fig. \ref{fig:par1}.

Similar analyses of 
the $2\omega$- and $\omega$- resonant contributions in the SHG 
have been also performed for the other SiC polytypes.
Comparisons between the z- polarized component
of the dielectric functions (DF) and the 333 components of the SHG 
for the 2H, 4H, 15R, and 6H polytypes
are shown
in Figs. \ref{fig:pol_2w} and \ref{fig:pol_w}.
These components involve the same (z- polarized)
momentum matrix element. This guarantees
that the selection rules for a given electronic transition will
be the same for both curves.
Again, all the transitions 
(including those with rather small oscillator strength)
are seen in both the DF and SHG curves, and
the $2\omega$- part of the SHG 
dominates in the whole frequency range considered in Fig. \ref{fig:pol_2w}.
The imaginary parts of the DF's look rather similar for all the polytypes
and only differ one from another in some fine details,
except for the pure hexagonal polytype, 2H.
However, the SHG's contain many more
features, and their changes from polytype to polytype are much more
dramatic.
This reflects the fact that 
the electronic structure for the higher polytypes becomes 
more and more complicated with increasing n.
This results in part from the above mentioned fact that the
number of bands increases because of 
the effects of the band foldings
(for layers with cubic stackings) and also from the appearance of
the new local symmetries (for hexagonal stackings).
Concomitantly, 
the analysis of the band-to-band contributions 
becomes increasingly complicated.
For the pure hexagonal polytype 2H, however,
we see that the 333 component consists of 
two wide strips, 
one between 3 eV and 4.3 eV and the other between
4.3 eV and 5.2 eV. The former results from
transitions from the two upper
valence bands to the first and second conduction bands
and the latter from transitions to the third and fourth
conduction bands (see Fig. \ref{fig:par1}).
The matrix element factor
for the first strip is negative while that for the second one is positive.
This is opposite to the case of 3C-SiC where 
the lower
energy peak is positive.
This large negative feature 
of the 333 component near the threshold
$E_g^d/2$ is unique to the 2H polytype, as was mentioned before.

Figure \ref{fig:real} shows the absolute value of the frequency
dependent 333 component of $\chi^{(2)}$ for the four non-cubic polytypes.
The reported measurements of frequency dependent results 
for other materials have generally been only for
the absolute values of the SHG rather than the real
or imaginary parts. Consequently, the numerical calculations of
this function is of interest for future comparisons to experimental data.
It can be seen that the lineshape
of the $|\chi^{(2)}(-2\omega,\omega,\omega)|$ for the pure hexagonal 2H
material again exhibits the greatest differences from those 
of the other polytypes.
Another interesing feature of $|\chi^{(2)}(-2\omega,\omega,\omega)|$ for 2H is
its very weak dispersion in the energy interval
between zero and $E_g^d/2$; that is, 
its coefficient of the $\omega^2$ term,
which describes the behavior in this region, is much smaller than
for the other polytypes.
For 4H, 15R, and 6H, the peak between 3 and 4 eV dominates.
Its position gradually increases from 3.4 eV
for 4H to 3.7 eV for 6H
and to 4.1 eV for 3C. 
This structure arises from the $2\omega$- resonant transitions
(see also Fig. \ref{fig:pol_2w}) and is common for all the polytypes
except 2H.
%These features can be traced back to direct transitions along the
%$\Gamma$-K line and along K-H where the
%top valence and lowest conduction bands are nearly parallel for all
%the hexagonal polytypes (see the discussion about the contributions
%of these transitions to the DF in Ref.\onlinecite{Lambrecht2}).

\subsection{Near band-edge fine structure}\label{finestruc}
An interesting feature of the various SHG spectra 
displayed in the previous sections  is the appearance of considerable
fine structure in the energy region between the threshold for the direct
interband transitions and the dominant peak around 3.5 eV.
These structures if detectable could provide important
information about the conduction and valence bands edges.
Neither the presently available 
reflectivity spectra,\cite{Lambrecht2} nor the
measurements of the dielectric function by spectroscopic ellipsometry 
\cite{Logothetidis} had the required resolution to provide  
such detailed information about this region of the spectra.

However, that information could be quite useful for 
determining the nature of the band edges and hence, for understanding
of the conduction properties in doped SiC polytypes. 
Information about
the band edges has previously been obtained by different methods.
These include analyses of the optically detected cyclotron
resonances \cite{LamSeg} and of the phonon replica spectra
associated with donor-bound
excitons. The latter are due to the phonons 
which have wave vector equal to the {\bf k} point location
of the conduction band minimum.\cite{Patrick,Choyke}
These analyses led to the conclusion that the minimum of the
conduction band in 4H is at the M point of the BZ in agreement with
the predictions of the band calculations.
For 6H the situation
is more complicated because not all of the multitude of phonon lines 
have as yet been resolved and also
because of the believed shallow nature of minimum itself.
In another experiment, the optical transitions from
the lowest conduction band to the higher bands 
were observed in fairly heavily 
n-type doped samples.\cite{Biedermann}
These spectra were recently analyzed by Lambrecht et al..\cite{Sukit}

The direct measurements of the SHG in this energy region
could provide additional information about the bands in the 
first few eV from the band edges.
Figure \ref{fig:par2} shows the analysis of the band-to-band
contribution to the $2\omega$- term in the imaginary
part of SHG (333 component)
for 4H and 6H polytypes. 
We analyze Im$\{\chi^{(2)}(-2\omega,\omega,\omega)\}$; but, 
of course, the same features would
appear in the $|\chi^{(2)}(-2\omega,\omega,\omega)|$.
For both polytypes the behavior of the SHG in this region is
determined by the transitions near the M- point (see Fig. \ref{fig:band}).
For 4H, the structure of Im$\{\chi^{(2)}(-2\omega,\omega,\omega)\}$ 
between 2.5 and 3 eV
results from transitions between the valence band $v2$ 
and the two lowest
conduction bands ($c1$ and $c2$).
The transitions to the first conduction band produce a positive
contribution to the SHG while those to 
the second band are negative. The separation between the broad maximum of SHG
around 2.7 eV (the {\bf a}- feature) and the minimum at 2.9 eV
(the {\bf b}- feature) is thus closely related to the 
energy difference between the two lowest conduction bands 
and the dispersion of the second highest valence band along the ML-line
(Fig. \ref{fig:band}).  Although the relation is not simple, 
it illustrates that
we can relate the features of the SHG curve to specific 
band-to-band contributions.

The situation in 6H is more complicated.
The total SHG curve in this region exhibits more structure than for 4H.
The lower peak at 2.38 eV (the {\bf a}- feature) 
results mainly from transitions
between the two valence bands $v2$ and $v3$ and the lowest conduction band
$c1$ (arrow 1 in the right hand panel of Fig. \ref{fig:band}). 
The partial contribution from these transitions 
decreases and changes sign
at higher energy and ultimately produces
a sharp minimum at 3.25 eV. This behavior is
also reflected in the total curve.
The second ({\bf b}- ) structure at 2.51 eV results from the
superposition of the first curve and the contribution from 
transitions from the same
valence bands ($v2$ and $v3$) to the second conduction band $c2$
(arrow 2).
The broad hump which starts at 2.75 eV originates from transitions
between the rather low-lying valence band $v5$ and the second
conduction band $c2$ (arrow 3). The last interesting feature
to be discussed is the hump
starting at 2.95 eV (the {\bf c}- feature).
This structure is associated with transitions
between the two upper valence bands ($v1$ and $v2$) and the third conduction
band $c3$ (arrow 4). 
While the total curve is fairly complicated, we see that energy differences 
in its features can be approximately related to half the differences
between the interband transition energies (Fig. \ref{fig:band}).

\section{Conclusions} \label{sec:con}

The second harmonic generation coefficients for five important
SiC polytypes, 3C, 6H, 15R, 4H, and 2H, have been calculated
over the energy range from zero to 13 eV. 
The results for the zero-frequency limit of the ratios of the 333 and 311
components 
are in good agreement with those obtained in pseudopotential
calculations,\cite{Chen1}
while the magnitudes of the individual components are found to be smaller
than those reported earlier. 
The polytypes studied span the full range of hexagonality
from zero for 3C to unity (complete hexagonality) for 2H SiC.
This allowed for a study of the trends in the SHG with this
quantity which characterizes the polytypes. 
It is found that with the exception of the 333 component of 2H-SiC, 
the frequency dependent SHG coefficients
in the different polytypes 
look rather similar over a broad frequency range.
The spectra gradually approach those for the zincblende
3C polytype but exhibit increasing complexity with increasing
length of the basic repeat unit of the layer stackings.
All independent tensor components were obtained, 
(and their symmetry relations verified), including the 
small ones which are only nonvanishing in the rhombohedral case.
We also find that the components which  only strictly equal 
in the static limit by the Kleinman symmetry still 
show great similarity over the full frequency range.
The spectral features in the 333 component were further 
analyzed by examining separately the $\omega$ and $2\omega$ resonances,
the absolute values and their correspondence to features in the 
33 component of the imaginary part of the dielectric function.
These one-to-one correspondences and further explicit
decompositions allowed us to 
assign the spectral features to specific band-to-band
contributions. We carried this out in considerable detail for the 
transitions in 4H and 6H in the low energy region which is the 
most accessible region experimentally. 

An important general conclusion of this last aspect of our work is 
that the frequency dependent $\chi^{(2)}$
may provide a useful tool for studying the electronic structure
especially for the wide band gap semiconductors. This is particularly
important for studying the bands in the vicinity of the band edges.
To our knowledge this has not been done before. 
The observation of structure in the linear response functions
arising from direct transitions involving these band edges is hindered by
the weakness of those features.
They probably can be seen in derivative spectra of the DF, i.e.
by a modulation spectroscopy.
In some sense, SHG provides information similar
to that obtained from the modulation spectroscopies.
For the SiC polytypes (and other wide band gap semiconductors as well)
the measurements of the 2$\omega$- part of the SHG involving 
the near band edges lie in the visible light region
where available lasers can be employed.
In spite of the fact that the amplitudes of the features
to be detected are rather small, they are comparable to 
zero frequency limits of the SHG which have been measured.
This implies that they can be experimentally detected.

\acknowledgements

Part of the computations were performed at the Ohio Supercomputer Center.
This work was supported by NSF (DMR95-29376).

%%%%%%%%%%%%%%%%%%%%%%%%%%%%%%%%%%%%%%%%%%%%%%%%%%%%%%%%%%%%%%%%%%%%%%%%
%
%       REFERENCES
%
%%%%%%%%%%%%%%%%%%%%%%%%%%%%%%%%%%%%%%%%%%%%%%%%%%%%%%%%%%%%%%%%%%%%%%%%

\newpage

\begin{table}
\caption{
LDA and LDA + ``scissors'' (with 1 eV shift) calculations
of the $\chi^{(2)}_{333}(0)$ and $\chi^{(2)}_{311}(0)$ 
compared with pseudopotential LDA calculations (in pm/V).
The sign of the SHG's from Ref. \protect\onlinecite{Chen1} 
was adapted to the present choice of coordinate system (see the text).
}
\begin{tabular}{lccccc}
$       $ & 2H    & 4H     &  15R  &  6H    & $\infty$H (3C) \\
\tableline
LDA       &       &        &        &      &          \\
$\chi^{(2)}_{333}(0)$ & 3.6  &  14.5 & 16.4 &  17.8 & 20.2  \\
$\chi^{(2)}_{311}(0)$ & -6.1 &  -8.9 & -9.3 &  -9.7 & -10.1 \\
\tableline
LDA + ``scissors'' &       &         &    &    &     \\
$\chi^{(2)}_{333}(0)$ &  2.5  &  9.2 & 10.4 &  11.4 & 13.0   \\
$\chi^{(2)}_{311}(0)$ & -3.9  & -5.7 & -6.1 &  -6.3 & -6.5   \\
\tableline
LDA (Ref. \protect\onlinecite{Chen1}) &      &      &     &    &      \\
$\chi^{(2)}_{333}(0)$ &   8.6 &  23.2  &  -  &   27.6  &   28.2  \\
$\chi^{(2)}_{311}(0)$ & -13.2 & -14.8  &  -  &  -15.0  &  -14.2  \\
\tableline
LDA + ``scissors'' (Ref. \protect\onlinecite{Chen1}) 
 &      &      &     &    &      \\
$\chi^{(2)}_{333}(0)$ &   5.8 & 15.6  &  -  &  18.6   &  18.4   \\
$\chi^{(2)}_{311}(0)$ & -8.8 & -10.0  &  -  &  -10.4  &  -9.2  \\
\tableline            
Expt. \tablenote{Rescaled from Ref. \onlinecite{Singh}, 
for undetermined $\alpha$-SiC polytype, see text.} &&&&& \\
$\chi^{(2)}_{333}(0)$ &   \multicolumn{4}{c}{$\pm23$} & - \\
$\chi^{(2)}_{311}(0)$ &   \multicolumn{4}{c}{$\mp14$} & -  \\
\end{tabular}

\label{tab:chi0}
\end{table}

\begin{figure}
%\begin{center}
\caption{
The calculated ratio
$-\chi^{(2)}_{333}(0)/\chi^{(2)}_{311}(0)$ 
as a function of hexagonality. 
The results of the present LDA calculations are shown as filled diamonds,
and those of Ref. \protect\onlinecite{Chen1} as unfilled diamonds.
}
\label{fig:ratio}
%\end{center}
\end{figure}

\begin{figure}
%\begin{center}
\caption{
The calculated values of the imaginary part of
$\chi^{(2)}(-2\omega,\omega,\omega)$ for the SiC polytypes considered:
a) 333 component; b) 311 component, c) 113 component.
}
\label{fig:chi_pol}
%\end{center}
\end{figure}

\begin{figure}
%\begin{center}
\caption{Ratio  $|\chi^{(2)}_{333}(-2\omega,\omega,\omega)|
/|\chi^{(2)}_{311}(-2\omega,\omega,\omega)|$ for the 
non-cubic SiC polytypes considered.}\label{figratiofreq}
%\end{center}
\end{figure}

\begin{figure}
%\begin{center}
\caption{
The calculated 222 (solid line) and 211 (dotted line)
components of the imaginary part of the SHG for the rhombohedral
15R polytype. 
The mirror plane is perpendicular to the x- axis.
}
\label{fig:trig}
%\end{center}
\end{figure}

\begin{figure}
%\begin{center}
\caption{
Upper part -
the analysis of the band-to-band contribution to the 123 component of the SHG
for 3C-SiC: the total SHG (solid line); upper valence band to first
conduction band contribution (dotted line); upper valence band to second
conduction band contribution (dashed line).
Lower part - the same for 2H polytype (333 component):
total SHG (solid line); 2 upper valence bands to first and second 
conduction bands contribution (dotted line); 2 upper valence bands to
third and fourth conduction bands contribution (dashed line).
}
\label{fig:par1}
%\end{center}
\end{figure}

\begin{figure}
%\begin{center}
\caption{
The values of the $2\omega$- resonant contributions
in the 333 component of SHG for the noncubic SiC polytypes 
considered (solid lines), 
in the units 10$^{-6}$ esu, compared
with the corresponding  
$\varepsilon_2^{zz}(2\omega)/50$
(dotted lines).
}
\label{fig:pol_2w}
%\end{center}
\end{figure}

\begin{figure}
%\begin{center}
\caption{
The values of the $\omega$- resonant contributions
in the 333 component of SHG for the noncubic SiC polytypes 
considered (solid lines), 
in the units 10$^{-6}$ esu, compared
with the corresponding 
$\varepsilon_2^{zz}(\omega)/100$
(dotted lines).
}
\label{fig:pol_w}
%\end{center}
\end{figure}

\begin{figure}
%\begin{center}
\caption{
The absolute values of the frequency dependent 333 component of
the SHG for the SiC polytypes considered. 
}
\label{fig:real}
%\end{center}
\end{figure}

\begin{figure}
%\begin{center}
\caption{
Analysis of the band-to-band contributions to 
Im$\{\chi^{(2)}_{333}(-2\omega,\omega,\omega)\}$
in the energy region associated with the direct transitions
between the near ``band edges'' for the 4H and 6H polytypes.
The total are given by
the thick solid lines; and, the numbers
labeling the curves for
the partial band-to-band contribution correspond to the interband 
transitions indicated in Fig. \ref{fig:band}.
}
\label{fig:par2}
%\end{center}
\end{figure}

\begin{figure}
%\begin{center}
\caption{
The electronic band structures for the 4H and 6H polytypes along 
the M-L line of the BZ. 
The LDA 
conduction bands are shifted up by a 1 eV ``correction''.
The transitions associated with the important contributions
to the SHG (and displayed in Fig. \ref{fig:par2}) are
indicated by arrows.
}
\label{fig:band}
%\end{center}
\end{figure}

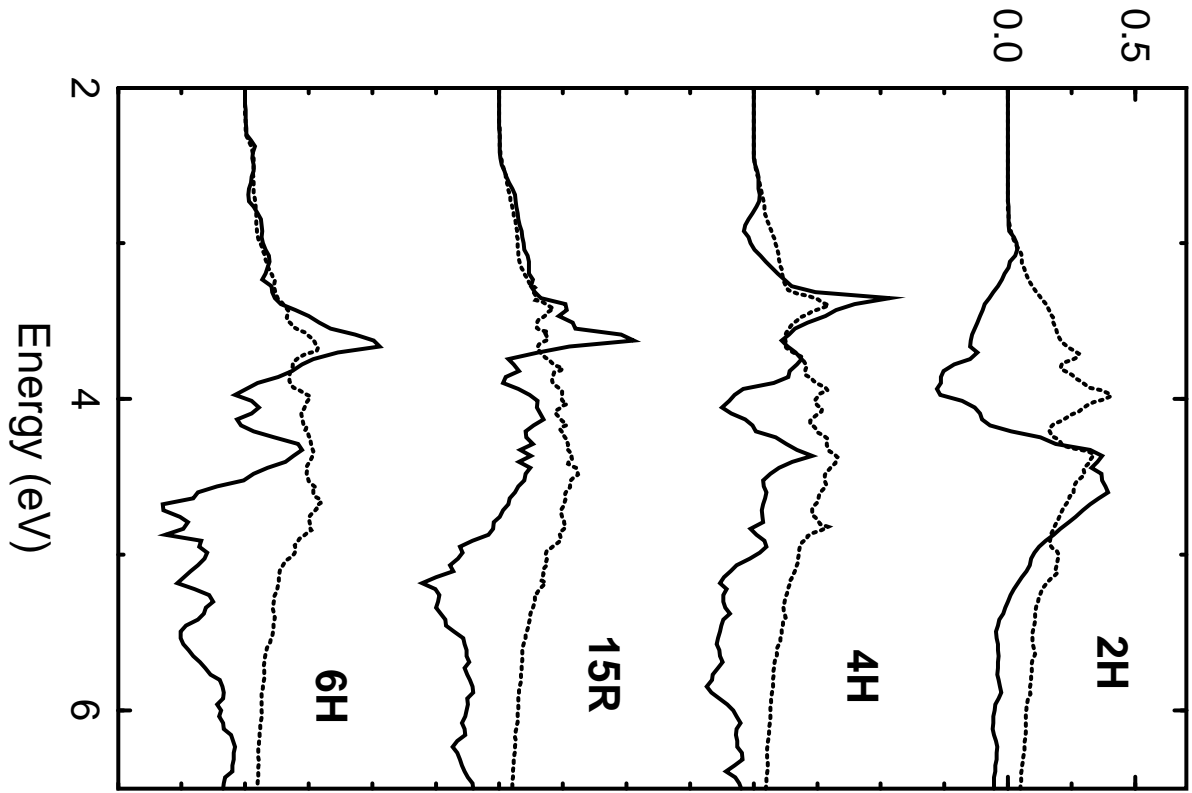
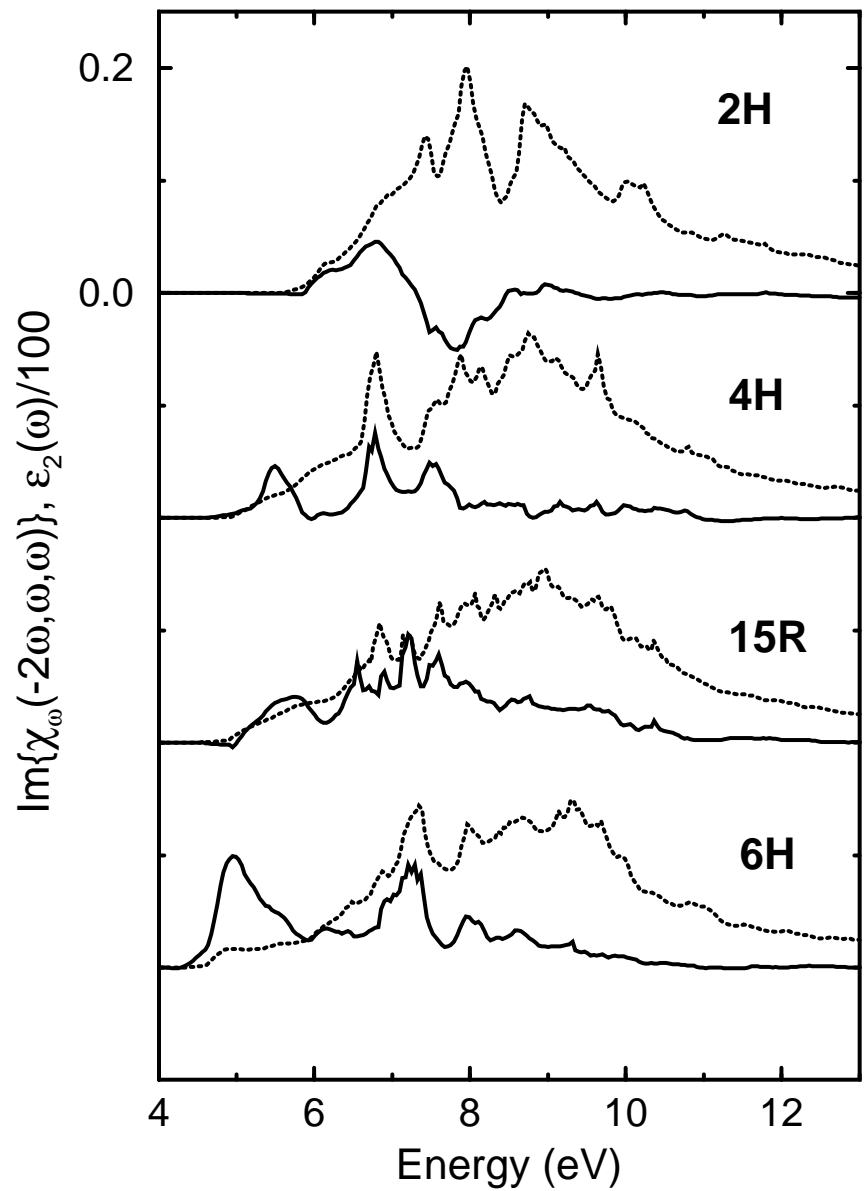
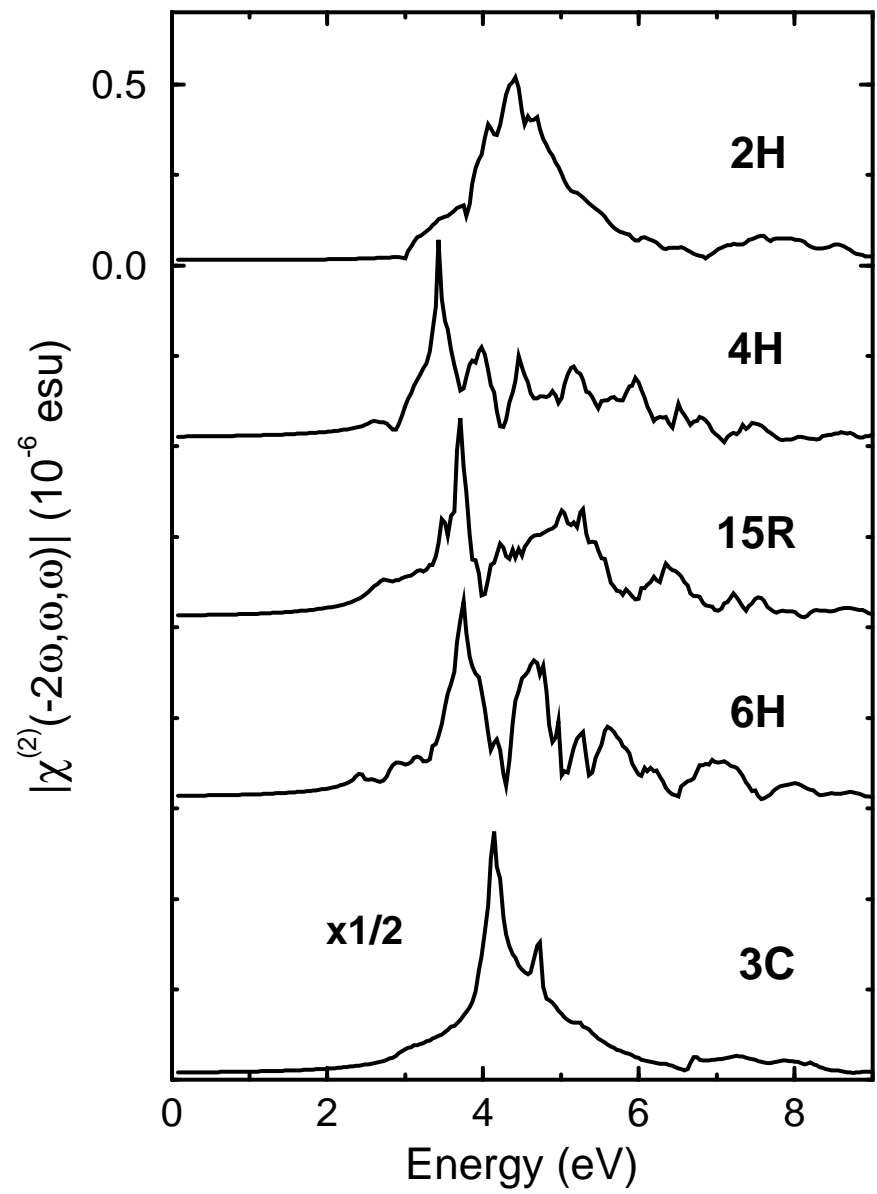
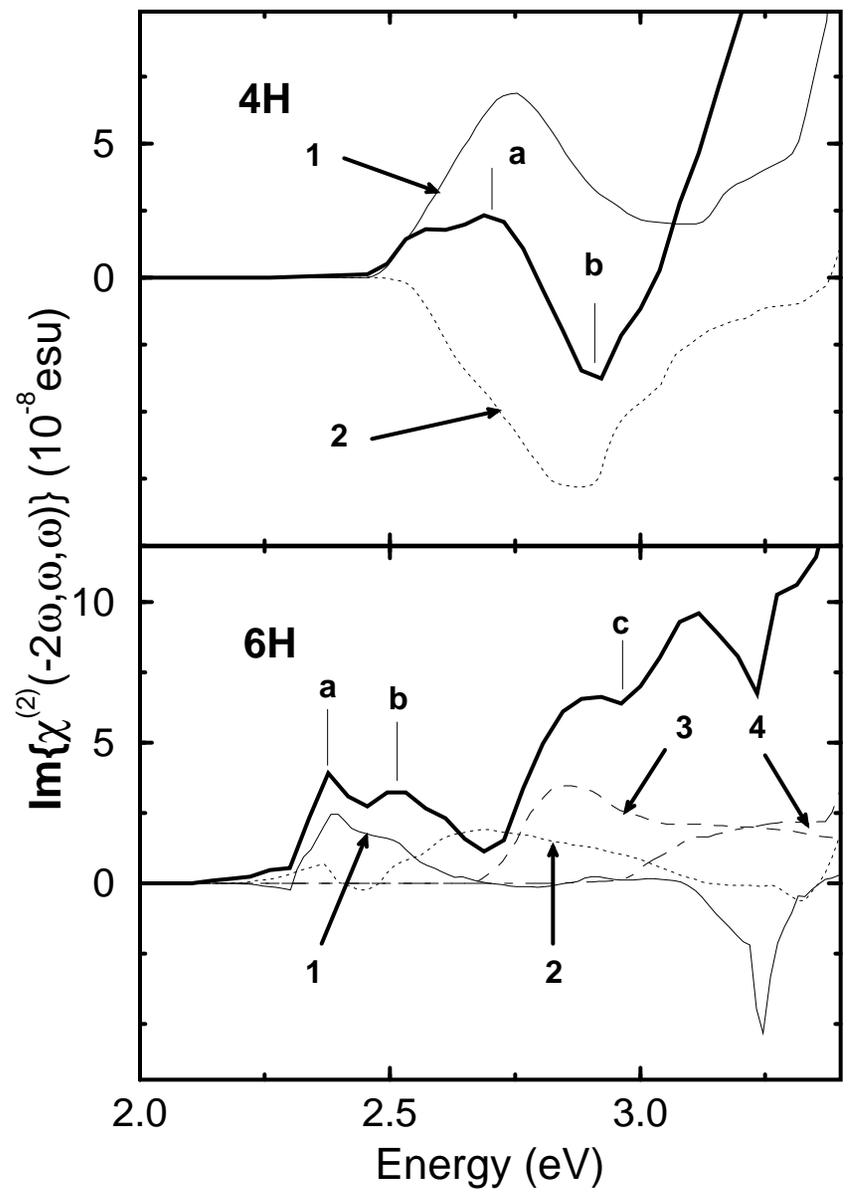
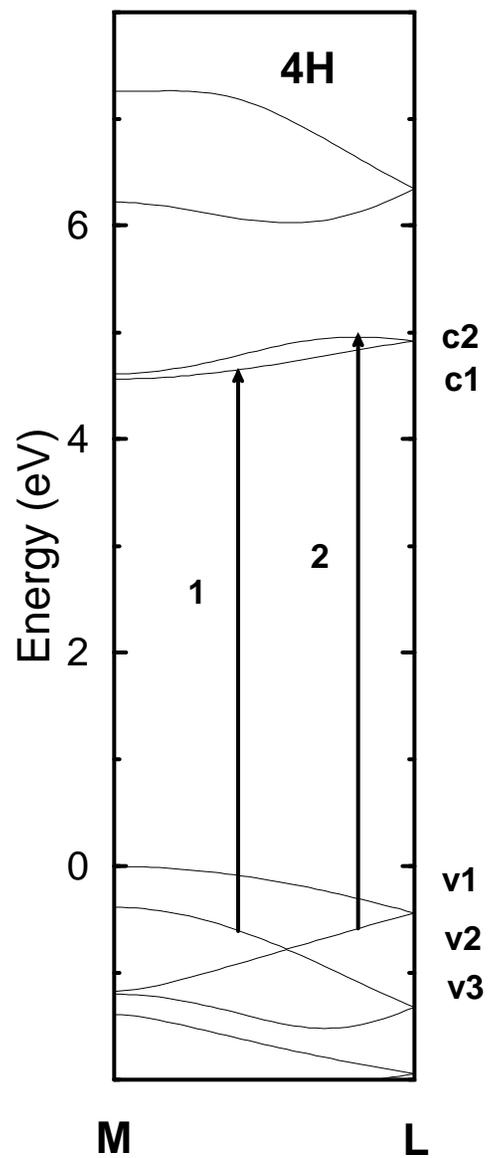
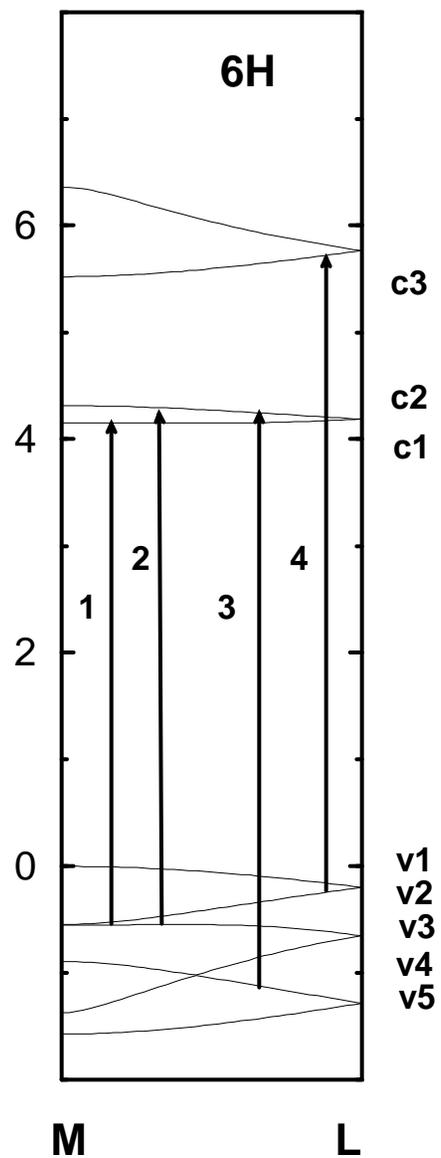
\end{document}